\documentclass[aps,twocolumn]{revtex4}
\usepackage{epsf}
\begin{document}
   
\title{Comment on `{\em Evidence for pairing above $T_{c}$ from the dispersion\\
        in the pseudogap phase of cuprates}' by A.\ Kanigel et al}
 
\author{T.\ Doma\'nski$^{1}$ and J.\ Ranninger$^{2}$}

\affiliation{$^{1}$
            Institute of Physics, Marie Curie-Sk\l odowska University, 
            20-031 Lublin, Poland \\
 	   $^{2}$
 	   Institut Neel CNRS, Dept.\ Matiere Condensee - Basses Temperatures,
           38042 Grenoble Cedex 09, France}

\begin{abstract}
In a recent preprint [cond-mat/0803.3052] A.\ Kanigel {\em et al} 
report evidence for Bogoliubov-type excitations in the pseudogap phase
in the anti-nodal region, where a robust  pseudogap remains well above
$T_{c}$. This important experimental result has been theoretically 
predicted by us almost 6 years ago on a basis of the phenomenological 
boson fermion model. An earlier theoretical prediction on the basis 
of this model was that of a pseudogap in the electron DOS,  setting 
in at some temperature $T^*$ and evolving into the superconducting 
gap upon approaching $T_c$. A natural logical pursuit of this early 
work was to show that, in order to have a superconducting state evolved 
out of a pseudogap state, the diamagnetic bosonic pair fluctuations 
(characterizing the pseudogap phase) have to be propagating modes 
and should be phase correlated over some finite distances above $T_c$. 
If so, then the pseudogap feature has to be reflected in characteristic 
features of the single particle excitations, showing remnants of 
the Bogoliubov modes inherent in the superconducting phase. Such  
Bogoliubov modes result from the dynamical feedback effects between 
single electron excitations and dynamical local pairing fluctuations. 
We briefly recollect here our theoretical results and confront them  
with the recent experimental findings.
\end{abstract}

\maketitle
 
\begin{figure}
\centerline{\epsfxsize=5cm \epsffile{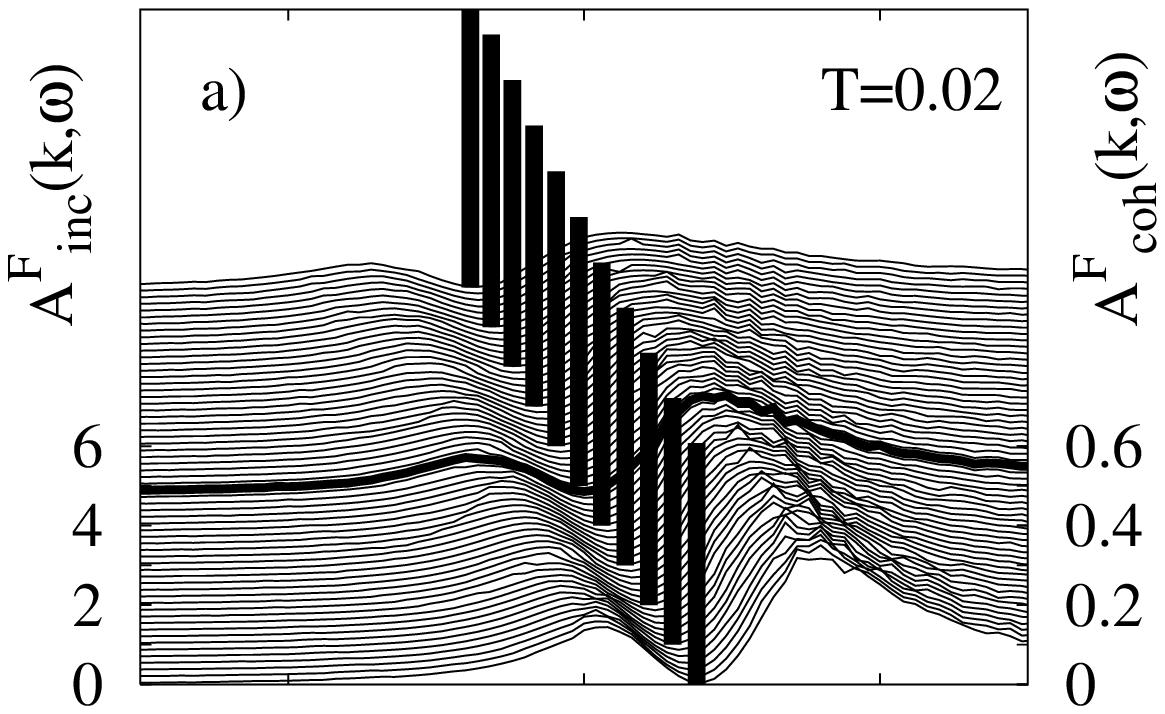}}
\vspace{-6mm}
\centerline{\epsfxsize=5cm \epsffile{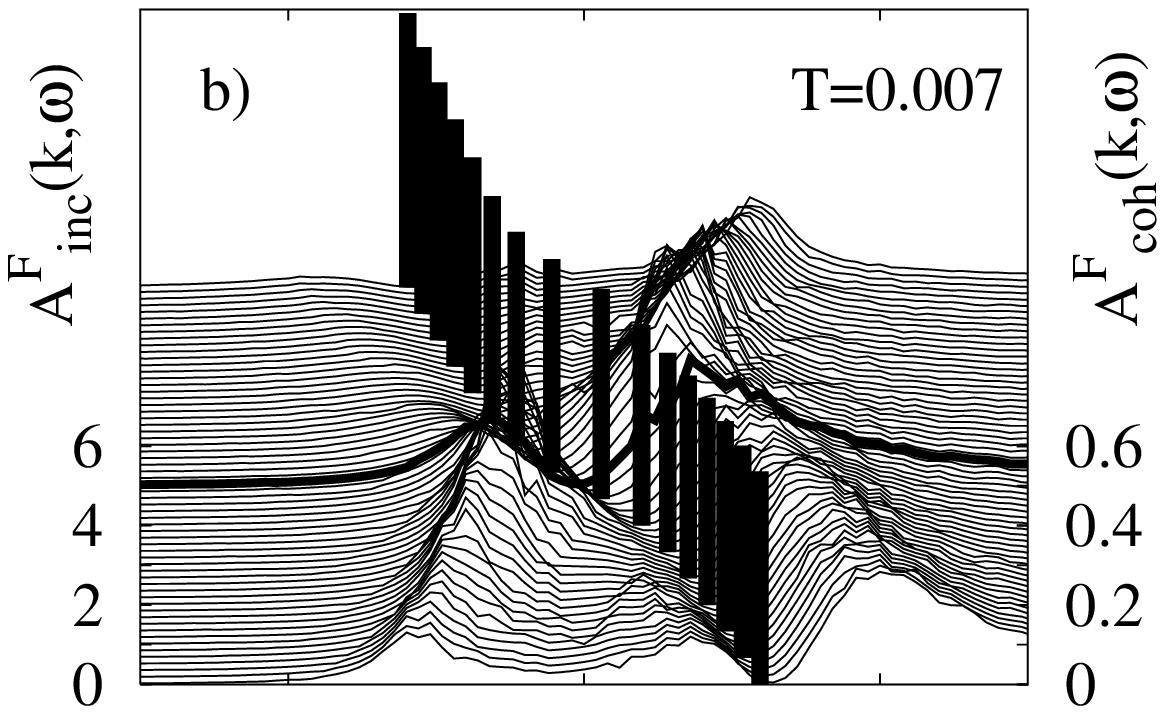}}
\vspace{-6mm}
\centerline{\epsfxsize=5cm \epsffile{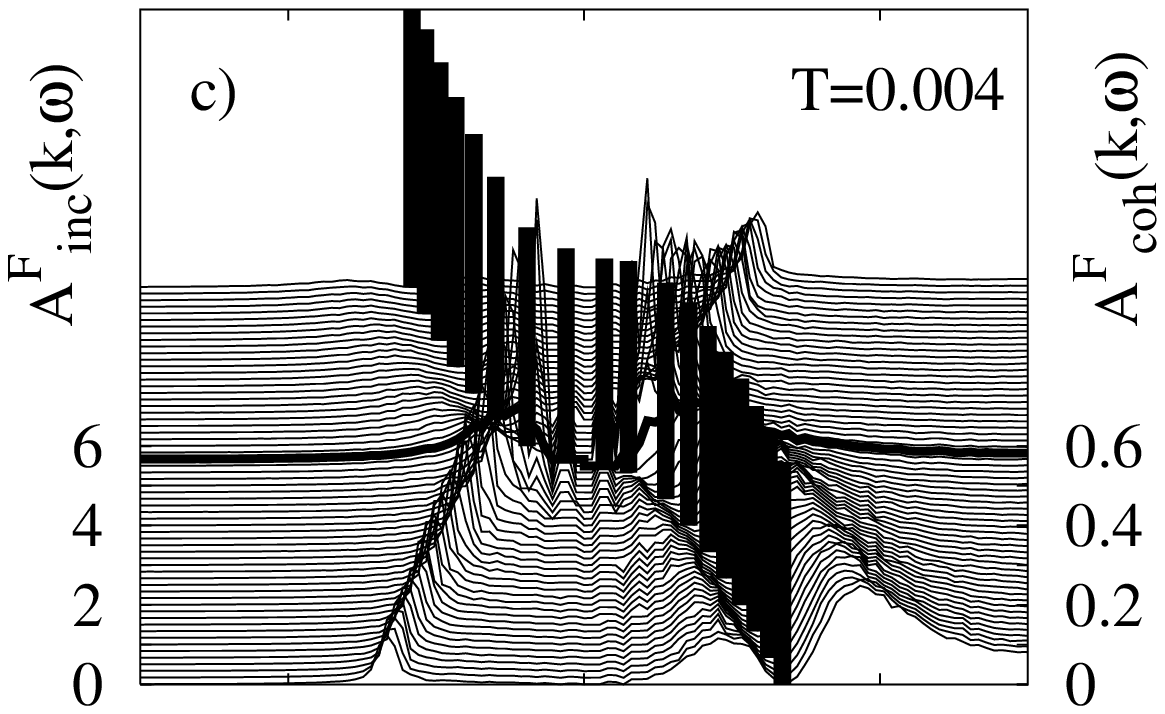}}
\vspace{-6mm}
\centerline{\epsfxsize=5cm \epsffile{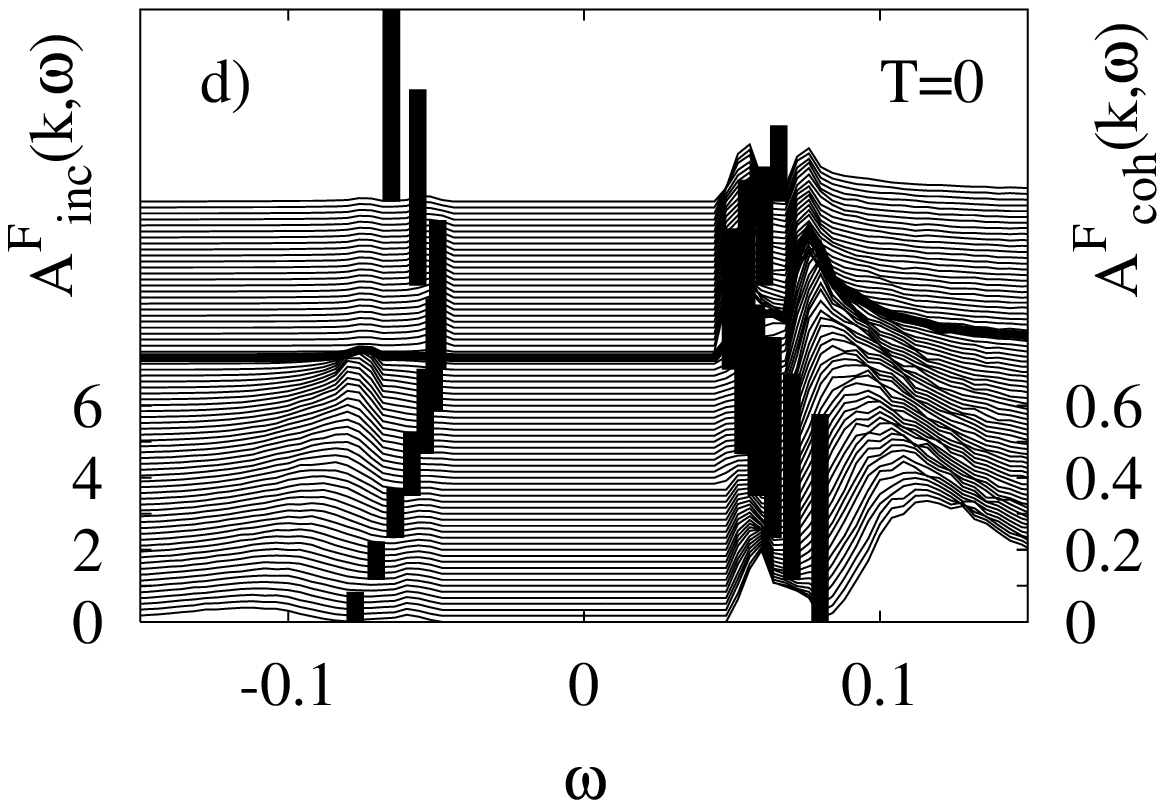}}
\caption{ {\footnotesize \sl
The single particle fermion spectral function $A^{F}({\bf k},\omega)$ 
decomposed into its coherent component $A^{F}_{coh}({\bf k},\omega)$ 
(thick bars whose height indicate the intensity of the delta like 
contributions) sitting on top of an incoherent component 
$A^{F}_{inc}({\bf k},\omega)$ in the vicinity of ${\bf k_{F}}$
indicated by the bold spectral line) for the normal phase 
(a) above $T^{*}$ ($T=0.02\;D$), (b) and (c)for the pseudogap 
region $T^{*}>T>T_{c}$ ($0.007\;D$, $0.004\;D$) and (d) for 
the superconducting phase (in the ground state $T=0$). 
The distance between the neighboring lines corresponds to 
changes in wavevector by multiples of $\Delta k = \pi /1000a$.
Figure is reproduced from our paper \cite{Domanski-03}.}}
\label{Fig1}
\end{figure}
%%%%%%%%%%%%%%%%%%%%%%%%%%%%%%%%%%%%%%%%%%%%%%%%%%%%%%%%%%%%%%%%%%%%%%%%%%

The mechanism of high temperature superconductivity (HTSC) is widely 
believed to be related to the strong correlations between electrons in 
the 2-dimensional CuO$_{2}$ planes. However, no specific microscopic 
model has been so far fully accepted, mainly because of conflicting 
interpretations of the pseudogap state in the underdoped cuprates. 
The recent experimental data \cite{Kanigel-08} unambiguously confirm 
that the pseudogap is a signature of pairing fluctuations. For the regions
in the Brillouin zone, where  the pseudo gap is present above $T_{c}$, 
these authors have indeed detected such a Bogoliubov-type excitation 
spectrum, as predicted by us theoretically \cite{Domanski-03}.

There are many theoretical approaches proposed to explain HTSC materials 
\cite{Review_papers}. Many of  them rely on assumption that $d$-wave 
superconductivity originates from the non-retarded intersite pairing. 
The corresponding two-body interactions are then transformed away via 
the usual Hubbard Stratonovich transformation, which introduces auxiliary 
bosonic pairing fields. Generally, many studies focus on the saddle point 
(mean-field) solution plus small (Gaussian) corrections around it. Such 
procedure is however questionable in the HTSC cuprates, where the fermion 
and boson degrees of freedom are strongly mixed with one another. 
 
We have for that reason preferred to follow a phenomenological approach 
assuming that along the antinodal directions the underlying physics can 
be described in terms of itinerant fermions hybridized with the local 
pairs via Andreev-type scattering. This, so-called Boson Fermion model 
was proposed well before the discovery of HTSC \cite{Ranninger-1985}. 
Already on a meanfield basis, this model showed the intricate interplay 
between pairing correlations and the opening of a gap in the single
particle DOS.
 
Upon going beyond the mean field, it clearly indicated a persisting 
pseudogap above $T_c$ \cite{Ranninger-1995}. This theoretical prediction 
was verified experimentally a year later or so by the Argonne and Stanford 
groups.

Following this initial theoretical work, it became clear that in order 
to proceed from the pseudogap into the superconducting phase upon lowering
the temperature, one required a selfconsistent approach, treating the pair
fluctuations and the single particle excitations on the same footing.  For
that to achieve we have used a numerical RG approach \cite{Wegner-94},  
which allowed us to account for a mutual renormalization  of the single 
and paired electrons via coupled flow equations \cite{Domanski-04}. From 
their solution we derived the resulting single particle spectral function 
$A({\bf k},\omega)$ \cite{Domanski-03}. For the energies $\omega<0$ (which
are probed by the direct photoemission), this spectral function for 
$T \leq T_c$ turned out to have the following form \cite{Domanski-04}

\begin{eqnarray}
A({\bf k},\omega\!<\!0) &=& | u_{\bf k}|^{2} 
\delta \left( \omega\!-\!\tilde{\xi}_{\bf k} \right) +
| v_{\bf k}|^{2}  \frac{\Gamma_{\bf k}/\pi} {(\omega 
+\!\tilde{\xi}_{{\bf k}})^{2}+\Gamma_{\bf k}^{2}}
\nonumber \\
&&+ \;A_{bg}({\bf k},\omega) \nonumber \\
\label{spectral} 
\end{eqnarray}
with qusiparticle dispersion $\tilde{\xi}_{\bf k}
=\sqrt{(\varepsilon_{k}-\mu)^{2}+\Delta_{{\bf k},pg}^2}$.
$\Gamma_{\bf k}$ denotes a broadening which increases with 
increasing temperature, while the pseudogap $\Delta_{{\bf k},pg}$
hardly changes, provided we are well below below $T^{*}$ $(T_c \leq T << T^*$). 
The remaining background $A_{bg}({\bf k},\omega)$ is rather rigid and
its contribution is not relevant to particle-hole mixing
arising from the pairing fluctuations.  Obviously for the
positive energies (measured by the inverse photoemission) the
spectral function becomes $A({\bf k},\omega>0) = | v_{\bf k}|^{2}
\delta ( \omega\!+\!\tilde{\xi}_{\bf k} ) + | u_{\bf k}|^{2}
\frac{\Gamma_{\bf k}/\pi} {(\omega -\!\tilde{\xi}_{{\bf k}})^{2}
+\Gamma_{\bf k}^{2}}+ A_{bg}({\bf k},\omega)$. 

\vspace{0.2cm}
In figure \ref{Fig1} we reproduce \cite{Domanski-03} the corresponding 
results obtained for the spectral function near the antinodal region
within the boson-fermion model scenario. One clearly notices the emergence
of the Bogoliubov-type spectrum both, below and above $T_{c}$. In the 
pseudogap phase, slightly above $T_c$, such Bogoliubov shadow modes 
appear broadened and such behavior is in agreement with the experimental 
findings reported by A.\ Kanigel et al \cite{Kanigel-08}. Upon further 
increasing the temperature those Bogoliubov shadow modes get overdamped 
and upon approaching $T^{*}$ they fade away, due to  life-time effects.  
A concomitant gradual closure of the pseudo-gap is then the signature 
of the phase uncorrelated pairing fluctuations. 
 
The presence of a Bogoliubov spectrum above $T_{c}$ together with 
other experimental facts, such as the residual diamagnetism (Ref.\ 
25 cited in \cite{Kanigel-08}) and the observation of vortices 
(Ref.\ 26 cited in \cite{Kanigel-08}) bring further evidence that 
$T_{c}$ must be related to a loss of long-range phase coherence. 
For the time-being, the experimental observation by A.\ Kanigel 
{\em et al} \cite{Kanigel-08} works strongly in favor of the 
precursor scenario, which initially has been proposed by one 
of us \cite{Ranninger-1988,Ranninger-1995} and well before 
the experimental verification of the pseudogap phase by ARPES
measurements.
 
We hope that in future, experimental groups would  give some credit 
to such theoretical predictions, which might have stimulated such 
beautiful and important experiments and the physical insights.


\begin{thebibliography}{11}
\bibitem{Kanigel-08}
     A.\ Kanigel, U.\ Chatterjee, M.\ Randeria, M.R.\ Norman, 
    G.\ Koren, K.\ Kadowaki, and J.C.\ Campuzano,
   cond-mat/0803.3052 (preprint).
\bibitem{Domanski-03}
     T.\ Doma\'nski and J.\ Ranninger, 
     Phys.\ Rev.\ Lett.\  {\bf 91}, 255301 (2003).
\bibitem{Review_papers}
     P.A.\ Lee, N.\ Nagaosa, and X.G.\ Wen,
     Rev.\ Mod.\ Phys.\ {\bf 78}, 17 (2006);
     C.M.\ Varma, Phys.\ Rev.\ B {\bf 73}, 155113 (2006);
     P.W.\ Anderson, P.A.\ Lee, M.\ Randeria, T.M.\ Rice 
     and F.C.\ Zhang, J.\ Phys.\ Condens.\ Matter 
     {\bf 16}, R755 (2005);
     M.\ Franz and Z.\ Tesanovic, 
     Phys.\ Rev.\ Lett.\ {bf 87}, 257003 (2001).
\bibitem{Ranninger-1985} 
    J.\ Ranninger, S.Robaszkiewicz, 
     Physica B{\bf 135}, 468 (1985).
\bibitem{Ranninger-1995}
     J.\ Ranninger, J.M.\ Robin, and M.\ Eschrig,
     Phys.\ Rev.\ Lett.\ {\bf 74}, 4027 (1995).
\bibitem{Wegner-94}
     F.\ Wegner, Ann.\ Physik {\bf 3}, 77 (1994);
     S.\ Kehrein, {\em Flow Equation Approach to Many Particle 
     Systems}, Springer Tracts in Mod.\ Phys.\ {\bf 217}, (2006).
\bibitem{Domanski-04}
     T.\ Doma\'nski and J.\ Ranninger, 
     Phys.\ Rev.\ B {\bf 70}, 184503 (2004);
%     T.\ Doma\'nski and J.\ Ranninger, 
     Phys.\ Rev.\ B  {\bf 63}, 134505 (2001).
\bibitem{Ranninger-1988}
    J.\ Ranninger, R.\ Micnas, and S.\ Robaszkiewicz, 
    Annales de Physique (France) {\bf 13}, 9410 (1988).
\end{thebibliography}
\end{document}